\documentclass[twocolumn,twoside,slac_two,superscriptaddress]{revtex4}
\usepackage{graphicx}
\usepackage{fancyhdr}
\usepackage{hyperref}
\hypersetup{
    colorlinks=true,
    urlcolor=magenta,
}

\usepackage{amsmath}
\newcommand{\galprop}{GALPROP}

\newcommand{\aap}{A\&A}

\newcommand{\fermilat}{\emph{Fermi}-LAT}

\begin{document}

\title{From Observations near the Earth to the Local Interstellar Spectra}

\author{S. {Della~Torre}}
\affiliation{INFN Milano-Bicocca, Piazza della Scienza 3, 20125 Milano, Italy}
\author{M. Gervasi}
\affiliation{INFN Milano-Bicocca, Piazza della Scienza 3, 20125 Milano, Italy}
\affiliation{Physics Department, University of Milano-Bicocca, Milano, Italy}
\author{D. Grandi}
\affiliation{INFN Milano-Bicocca, Piazza della Scienza 3, 20125 Milano, Italy}
\author{G. Johannesson}
\affiliation{Science Institute, University of Iceland, Dunhaga 3, IS-107 Reykjavik, Iceland}
\author{G. La~Vacca} 
\affiliation{INFN Milano-Bicocca, Piazza della Scienza 3, 20125 Milano, Italy}
\author{N. Masi}
\email[corresponding author:]{masin@bo.infn.it}
\affiliation{INFN, Bologna, Italy}
\author{I.V. Moskalenko} 
\affiliation{Hansen Experimental Physics Laboratory, Stanford University, Stanford, CA 94305}
\affiliation{Kavli Institute for Particle Astrophysics and Cosmology, Stanford University, Stanford, CA 94305}
\author{E. Orlando}
\affiliation{Hansen Experimental Physics Laboratory, Stanford University, Stanford, CA 94305}
\affiliation{Kavli Institute for Particle Astrophysics and Cosmology, Stanford University, Stanford, CA 94305}
\author{T.A. Porter}
\affiliation{Hansen Experimental Physics Laboratory, Stanford University, Stanford, CA 94305}
\author{L. Quadrani}
\affiliation{INFN, Bologna, Italy}
\affiliation{Physics Department, University of Bologna, Bologna, Italy}
\author{P.G. Rancoita}
\affiliation{INFN Milano-Bicocca, Piazza della Scienza 3, 20125 Milano, Italy}
\author{D. Rozza}   
\affiliation{INFN Milano-Bicocca, Piazza della Scienza 3, 20125 Milano, Italy}
\affiliation{Physics Department, University of Milano-Bicocca, Milano, Italy}

\begin{abstract}
Propagation of cosmic rays (CRs) from their sources to the observer is described mainly as plain diffusion at high energies, while at lower energies there are other physical processes involved, both in the interstellar space and in the heliosphere. The latter was a subject of considerable uncertainty until recently. New data obtained by several CR missions can be used to find the local interstellar spectra (LIS) of CR species that would significantly reduce the uncertainties associated with the heliospheric propagation.
In this paper we present the LIS of CR protons and helium outside the heliospheric boundary. The proposed LIS are tuned to accommodate both, the low energy CR spectra measured by Voyager 1, and the high energy observations publicly released by BESS, Pamela, AMS-01 and AMS-02. The proton and helium LIS are derived by combining CR propagation in the Galaxy, as described by GALPROP, with the heliospheric modulation computed using the HelMod Monte Carlo Tool. The proposed LIS are tuned to reproduce the modulated spectra for both, high and low, levels of solar activity.
\end{abstract}

\maketitle

\thispagestyle{fancy}

\section{Introduction}

\enlargethispage{\baselineskip}

In recent years, considerable advances in astrophysics of CRs have become possible due
to superior instrumentation launched into space and to the top of the atmosphere.
The launch of Payload for Antimatter Matter Exploration and Light-nuclei Astrophysics (PAMELA) in 2006 \citep{2007APh....27..296P}, followed by the {\it Fermi} Large Area Telescope (\fermilat) in 2008 \citep{2009ApJ...697.1071A}, and the Alpha Magnetic Spectrometer--02 (AMS--02) in 2011~\citep{2013PhRvL.110n1102A}
signify the beginning of a new era in astrophysics. Their excellent quality data are leading to breakthroughs in our understanding of CR sources, particle transport in the interstellar medium and in the heliosphere, as well as the details of the structure of the interstellar medium.

In this paper, the most recent CR measurements were used to unveil the true local interstellar spectra of CR protons and helium. Our approach combines two state-of-the-art propagation packages, \galprop{} for the Galaxy and HelMod for the heliosphere \citep{HelmodECRS2016}, into a single framework that is run to reproduce direct measurements of CR species at different modulation levels and at both polarities of the solar magnetic field. The proposed LIS accommodate both the low energy interstellar CR spectra measured by Voyager 1 and the higher energy observations by BESS, Pamela, AMS-01, and AMS-02. Here we are providing only illustrative results, more details could be found in a forthcoming paper.
 
\begin{table}[b!]
\begin{center}
\caption{Propagation parameters, obtained with the MCMC posterior distributions and the GALPROP-HelMod calibration\label{tbl-1}}
\begin{tabular}{ccccc}
\tableline\tableline
N & Parameters & Best Value & Units & Scan Range \\
\tableline
1 & $z$ &4.0 & kpc  &[1-10] \\
2 & $D_{0}/10^{28}$ &4.3 & cm$^{2}$ s$^{-1}$ &[1-10]\\
3 & $\delta$ &0.39 & \dots &[0.3-0.9]\\
4 & $V_{\rm Alf}$ &28.5 & km s$^{-1}$  &[0-40]\\
5 & $V_{\rm conv}$ &12.5 & km s$^{-1}$  &[0-20]\\
6 & $dV_{\rm conv}/dz$ &9.8 & km s$^{-1}$ kpc$^{-1}$ &[0-20]\\
\tableline
\end{tabular}
\end{center}
\end{table}

\section{Markov Chain Monte Carlo Approach to interstellar propagation }
\label{Sect::MCMC}

The MCMC interface to \galprop{} v55 was developed on the basis of CosRayMC~\citep{2012PhRvD..85d3507L} and, in general,
from the COSMOMC package \citep{2002PhRvD..66j3511L}, embedding the \galprop{} framework into the MCMC scheme.
Sampling of \galprop{} CR production and propagation parameters employs an iterative procedure while using AMS-02 data as observational constraints. This allows a considerable parameter space to be explored and provides estimates of the most probable values of the parameters together with their 68\%/95\% confidence intervals.

Six main propagation parameters (Table \ref{tbl-1}), that determine the overall shape of CR spectra,
were taken into account in the scan using the 2D \galprop{} model: the Galactic halo half-width $z$,
the normalization of the diffusion coefficient $D_0$ and the index of its rigidity dependence $\delta$,
the Alfv\'en velocity $V_{\rm Alf}$, the convection velocity and its gradient ($V_{\rm conv}$, $dV_{\rm conv}/dz$).
The single values of the injection indices were subsequently tuned together with HelMod model version 3.0 \citep{HelmodECRS2016}.


The solar modulation is made by means of the numerical functions, which provide the solution of the CR transport through the heliosphere and are embedded in the HelMod code. The high energy break, or \textit{change of slope}, for protons and helium highlighted by ATIC~\citep{2006AdSpR..37.1950A}, CREAM~\citep{2011ApJ...728..122Y}, PAMELA \citep{2011Sci...332...69A} and AMS-02 \citep{2015PhRvL.114q1103A,2015PhRvL.115u1101A} is firstly computed introducing \textit{ad hoc} an additional injection index $\gamma$ for nuclei.
This second break is tuned to be in agreement with CREAM-I data above AMS-02 range.
The experimental observables used in the MCMC scan include all published AMS-02 data on protons~\citep{2015PhRvL.114q1103A}, helium~\citep{2015PhRvL.115u1101A}, B/C ratio (preliminary results from \citep[][]{ICRC2015_BoC}), and electrons \citep{2014PhRvL.113l1102A}, while positrons and antiprotons were excluded. 

\begin{figure}[tb!]
\centerline{
\includegraphics[width=0.5\textwidth]{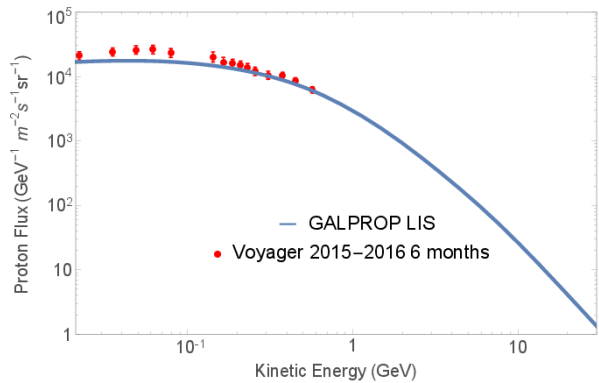}
}

\caption{A comparison of the DCR best set proton LIS (blue curve) with Voyager 1 2015-2016 monthly averaged data shown as a function of kinetic energy.}
\label{fig:Voyager_Proton}
\end{figure}

The results of the simulations show that simultaneous inclusion of diffusion, convection, and reacceleration
is required to reproduce AMS-02 measurements, while plain diffusion scenarios are excluded \citep{2007ARNPS..57..285S}.
Since $p$ and He spectra and the B/C ratio are not affected by the pulsar contribution nor by the Dark Matter annihilations,
they can be used to constrain the modulated spectra provided by HelMod making the results unbiased and easy to understand.
Therefore, uncertainties associated with the interstellar propagation can be significantly reduced relatively to other analyses
\citep{2001ApJ...555..585M,2002astro.ph.12111M,2006ASSL..339.....D,Stanev2010,2011JCAP...03..051C,2011ApJ...729..106T,2012A&A...539A..88C,2012A&A...544A..16T, Masi2013}, providing an order of magnitude improvement in the accuracy: in our analysis, the final errors associated with the determination of the major propagation parameters are reduced to $\sim$5--10\%.
Table \ref{tbl-1} lists the propagation parameters used in the scan, their prior ranges, and the computed best values.  



\section{Proton and Helium LIS outside Modulated Energy Region}\label{Sect::LISOutsideMod}


The direct measurements of proton and helium LIS are now available at both low and high energies. At low energies, measurements of CR fluxes are provided by Voyager 1 that crossed the Termination Shock (TS) in the second half of 2012 \citep{2013Sci...341..150S,2016ApJ...831...18C}. We took the latest Voyager 1 2015-2016 data averaged over monthly intervals (http://voyager.gsfc.nasa.gov/heliopause/vim/month\-ly/index.html).
The average of six months of Voyager 1 data for protons is shown in Figure~\ref{fig:Voyager_Proton}. The error associated with each data point is chosen conservatively to be equal to the variation of the monthly average, but not smaller than 15\% of the flux value. The model provides a good description of proton and helium LISs at low energies. We also emphasize that Voyager 1 data were not included into the MCMC scan, and a remarkable agreement between the model predictions and the LIS data supports our approach. The low energy LIS by Voyager 1 that are reproduced by \galprop{} are linked to the modulated AMS-02 data using the HelMod code.

\begin{figure}[tb!]
\centerline{
\includegraphics[width=0.49\textwidth]{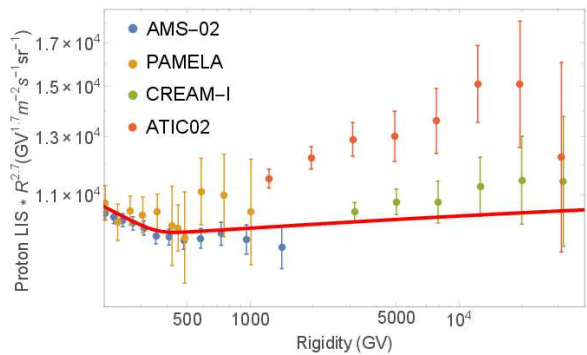}
}
\caption{The best fit proton LIS (red curve) is compared with high energy data by AMS-02, CREAM-I, ATIC-02, and PAMELA.
}
\label{fig:highp_lis}
\end{figure}


At high energies, where CR fluxes are not affected by the heliospheric modulation, we use AMS-02 data up to $\sim$2 TV and extend the rigidity range to 20--30 TV using data taken by CREAM-I and ATIC-02 (Figure ~\ref{fig:highp_lis}). In this energy range the data are scarce and there is a systematic discrepancy between CREAM-I and ATIC-02. Extrapolations of proton and He spectra by AMS-02, even not perfect, seem to prefer CREAM-I data. The relatively smaller error bars of CREAM-I vs.\ AMS-02 data at high energies drive the fit to result in flatter index values.

\begin{figure}[tb!]
\centerline{
 \includegraphics[width=0.47\textwidth]{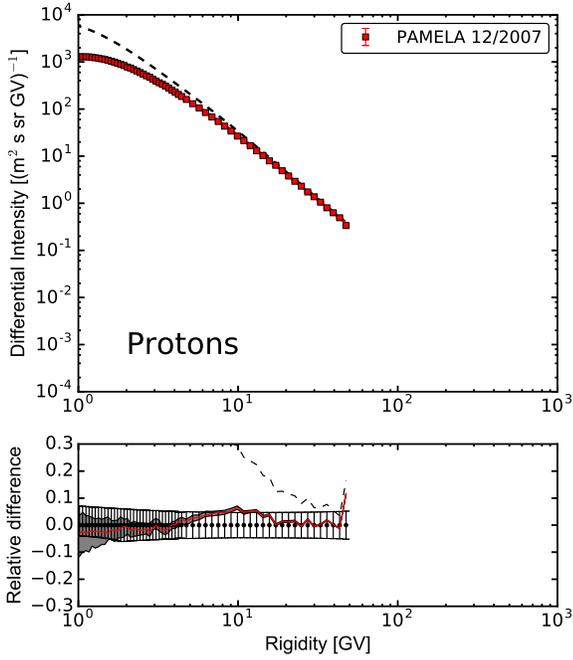}
 }
 \caption{Proton differential intensity for a low solar activity period (PAMELA 2007). Points represent experimental data, dashed line is the \galprop{} LIS, and solid line is the computed modulated spectrum. The bottom panel is the relative difference between the numerical solution and experimental data.}
 \label{fig:Proton_Low}
\end{figure}

\begin{figure}[tb!]
\centerline{
 \includegraphics[width=0.47\textwidth]{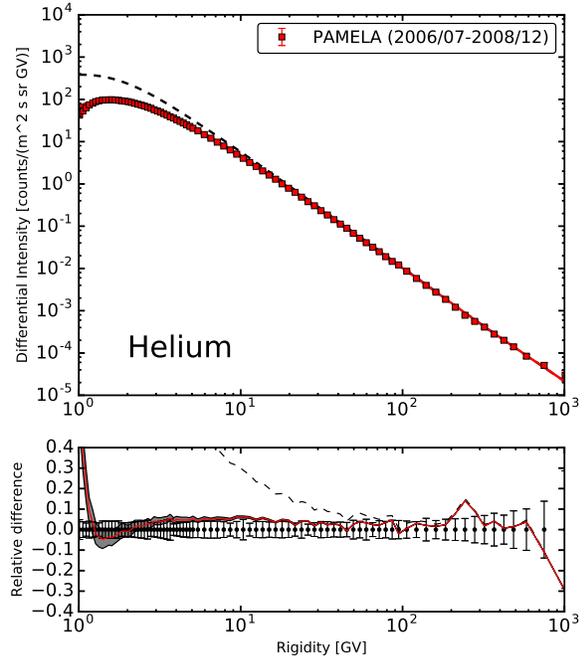}
 }
 \caption{Helium differential intensity for a low solar activity period (averaged PAMELA 2006-2009 measurements). 
 See Figure \ref{fig:Proton_Low} for legend.
}
 \label{fig:Helium_Low}
\end{figure}

We provide the analytical functional dependence of the derived LIS as a function of rigidity. To achive the required accuracy, especially in the AMS-02 range, the fit was split into two rigidity intervals, roughly below and above 1 GV. The search of the analytic solutions was guided by an advanced MCMC fitting procedure such as Eureqa (http://www.nutonian.com/products/eureqa). The combined LIS formula looks like:
\begin{align}\label{EQ::an}
&F(R)\times R^{2.7} = \\
&\left\{
\begin{array}{ll}
\sum_{i=0}^5 a_i R^{i}, &R\le1\ {\rm GV},\nonumber\smallskip\\
b + \frac{c}{R} + \frac{d_1}{d_2+R} + \frac{e_1}{e_2+R} + \frac{f_1}{f_2+R} + g R, & R>1 \ {\rm GV},\nonumber
\end{array}
\right.\nonumber
\end{align}
where $a_i, b, c, d_i, e_i, f_i, g$ are the numerical coefficients reported in Table~\ref{tbl-param}.
The derived expressions are (i) quite similar (at the level of $<$1--2\%) to numerical solutions in 5 orders of magnitude energy interval, including the spectral flattening at high energies, and (ii) are based on Voyager 1, AMS-02, and CREAM-I data.

\section{Data at Earth}\label{Sect::DataAtEarth}

As described in the previous sections, our approach combines two state-of-the-art codes, \galprop{} for interstellar propagation and HelMod for heliospheric propagation, within a single framework for the first time. AMS-02 data, which is guiding the refinement of the propagation scheme, is a vital ingredient of this approach. The converse is also true, the refined propagation scheme is beneficial for interpretation of the AMS-02 data. A comparison is made with observational data for conditions of low (i.e., 1997--1998, 2006--2010) and high solar activity (i.e., 2000--2002, 2011--2013), and then with the moderate activity period, thus providing an unique model that is valid for the entire solar cycle. 

\subsection{Low Solar Activity} \label{Sect::LowSolar}

During the solar minimum period, the heliospheric magnetic field (HMF) forms a regular structure thus requiring an inclusion of the magnetic drift effects; the latter is widely accepted in the literature, see e.g. \citep{Jokipii77,Potgieter85,BoellaEtAl2001,Strauss2011,DellaTorre2012,DellaTorre2013AdvAstro}. The solar minimal activity between cycles 23 and 24 was recently studied by the PAMELA instrument, e.g., \citep[][]{PamelaProt2013}. Such a period was characterized by a negative HMF polarity ($A<0$), that results in a more uniform latitudinal distribution in the inner part of the heliosphere for positively charged CR species. Examples of the proton and helium spectra during the low activity period are shown in Figures \ref{fig:Proton_Low} and \ref{fig:Helium_Low}.

\begin{figure}[tb!]
\centerline{
 \includegraphics[width=0.47\textwidth]{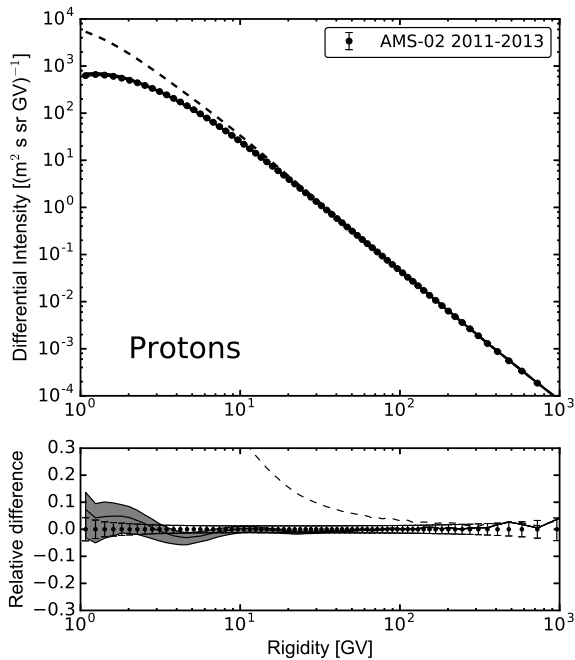}}
\caption{The differential intensities of CR protons measured by AMS-02 (high solar activity period). 
See Figure \ref{fig:Proton_Low} for legend.}
 \label{fig:p_High}
\end{figure}

\begin{figure}[tb!]
\centerline{
 \includegraphics[width=0.47\textwidth]{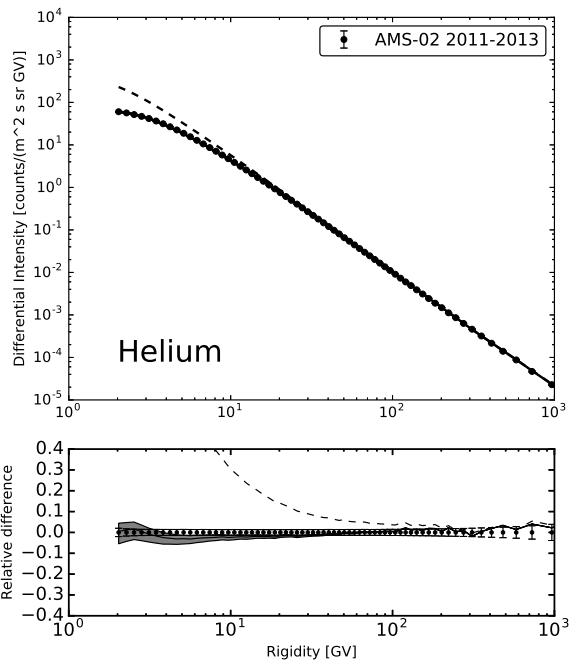}
 }
 \caption{The differential intensities of CR helium measured by AMS-02 (high solar activity period). 
 See Figure \ref{fig:Proton_Low} for legend.}
 \label{fig:He_High}
\end{figure}

\begin{table*}[bt!]
\centering
\caption{Parameters of the analytical fits to the proton and He LIS.\label{tbl-param}}
\begin{tabular}{|c|c|c|c|c|c|c|c|c|c|c|c|c|c|c|c|}
\hline
$$ & $a_0$ & $a_1$ & $a_2$ & $a_3$ & $a_4$ & $a_5$ & $b$ & $c$ & $d_1$ & $d_2$ & $e_1$ & $e_2$ & $f_1$ & $f_2$ & $g$ \\
\hline
$p$ & $94.1$ & $-831$ & $0$ & $16700$ & $-10200$ & $0$ & $10800$ & $8590$ & $-4230000$ & $3190$ & $274000$ & $17.4$ & $-39400$ & $0.464$ & $0$ \\
\hline
He & $1.14$ & $0$ & $-118$ & $578$ & $0$ & $-87$ & $3120$ & $-5530$ & $3370$ & $1.29$ & $134000$ & $88.5$ & $-1170000$ & $861$ & $0.03$ \\
\hline
\end{tabular}
\end{table*}%

\subsection{High Solar Activity}\label{Sect::HighSolar}

High solar activity periods are challenging from the viewpoint of theory of the heliospheric transport. The high frequency of solar events disturbs the interplanetary medium and disrupts the HMF that became difficult to model. An important change is the lack of regular structure of the HMF that completely suppresses the charge-sign dependence related to the magnetic drift process.
We introduced an additional correction factor that suppresses any drift velocity during the solar maximum.
AMS-02 provides an unique data-set integrated over 3 years \citep{2015PhRvL.114q1103A,2015PhRvL.115u1101A} of observations during the solar activity peak of cycle 24. The described model allows the average proton and helium spectra measured by AMS-02 during high solar activity periods (see Figures \ref{fig:p_High},\ref{fig:He_High}) to be well reproduced by the modulated ones from Helmod code.

\section{Conclusions}

Hundred years after the discovery of CRs, the unprecedented precision of AMS-02 instrument and its vast energy coverage promise solutions of many long-standing astrophysical puzzles. Once the spectra of all elements through iron measured with a few per cent accuracy up to several TV are released, they can be used to identify the sources of CRs and their propagation history, reveal the properties of the interstellar medium, and transform the whole field of astrophysical science.
The \galprop{}-HelMod framework is providing an example of a self-consistent and concise description of CR propagation from the Galactic scale down to the inner heliosphere. Elimination of the uncertainties in the astrophysical backgrounds would, in turn, enable us to search for traces of exotic physics.
The resulting LIS accommodate both the very low energy interstellar CR spectra measured by Voyager 1 and the higher energy observations at Earth publicly released by PAMELA and AMS-02.

\acknowledgements
This work is supported by ASI (Agenzia Spaziale Italiana) under contract ASI-INFN I/002/13/0 and and ESA (European Space Agency) contract 4000116146/16/NL/HK.
Igor Moskalenko, Elena Orlando, Troy Porter acknowledge support from NASA Grants Nos.~NNX13AC47G and NNX17AB48G, Elena Orlando additionally acknowledges support from NASA Grants Nos.~NNX16AF27G and NNX15AU79G, and Troy Porter additionally acknowledges support from NASA Grant No.~NNX10AE78G.

\bibliography{bibliography}
\end{document}